\documentclass[a4paper,reqno,oneside]{article}

\def\pf{\begin{proof}}
\def\pfk{\end{proof}}

\usepackage{amssymb,amsmath,amsthm}

\def\bp{\begin{proof}}
\def\ep{\end{proof}}

\theoremstyle{theorem}
\newtheorem{lem}{Lemma}[section]

\newtheorem{thm}[lem]{Theorem}

\theoremstyle{definition}
\newtheorem{de}[lem]{Definition}

\newtheorem{ex}{Example}

\begin{document}
\title{On representations of Lie algebras compatible with a grading}

\title{On representations of Lie algebras compatible with a grading}

\author{Miloslav Havl\'{\i}\v{c}ek,  Edita Pelantov\'a, Ji\v{r}\'{\i} Tolar
}

\maketitle

\begin{center}
{Doppler Institute, FNSPE, Czech Technical University in Prague,
Czech Republic
 }
\end{center}

\begin{abstract}

 The paper extends existing Lie algebra representation
theory related to Lie algebra gradings. The notion of a
representation compatible with a given grading is defined and
applied to finite-dimensional representations of
$sl(n,\mathbb{C})$ in relation with its $\mathbb{Z}_2$-gradings.
For representation theory of $sl(n,\mathbb{C})$ the
Gel'fand-Tseitlin method turned out very effective.
\end{abstract}


\section{Introduction}
Contractions of Lie algebras are usually defined inside the Lie
algebra structure. Remaining inside the framework of Lie algebras,
a slightly different notion of graded contractions was proposed in
\cite{MoPa}. Nevertheless, for physical applications only
representations of Lie algebras are relevant. In this matter the
only existing mathematical theory \cite{MuPa} is in a very
preliminary  shape. There has been a short note \cite{PaTo}, but
nobody has systematically investigated the representations of Lie
algebras related to their gradings. Our paper can be considered as
a starting point of such an investigation.

\subsection{ Graded contraction of Lie algebras}\label{zacatek}

\bigskip

A {\bf grading} of a Lie algebra $L$ is a decomposition $\Gamma$
of the vector space $L$ into vector subspaces $L_j$, $j \in
\mathcal J$, such that $L$ is a direct sum of these subspaces
$L_j$, and, for any pair of indices $j, k \in \mathcal J$, there
exists $l \in \mathcal J$ such that $\left[L_j, L_k \right]
\subseteq L_l$. We denote the grading by \[ \Gamma : L
=\bigoplus_{j \in \mathcal J} L_j. \] Clearly, for any Lie algebra
$L \neq \{0\}$, there exists the trivial grading $\Gamma : L = L$,
i.e. the Lie algebra is not split up at all. The opposite extreme
of splitting the Lie algebra into as many subspaces $L_j$ as
possible is called fine grading. Let us note that in our
definition of grading we do not exclude trivial subspaces
$L_i=\{0\}$. It follows directly from the definition that  for any
grading $\Gamma: \bigoplus_{j \in \mathcal J}
 L_j$ and any automorphism  $g \in {\mathcal {A}}ut\, L$ the
 decomposition $\Gamma':\bigoplus_{j \in \mathcal J}
 g(L_j)$ is a grading as well. The gradings $\Gamma$ and  $\Gamma'$
 are called equivalent.

Now we describe a specific type of a grading, namely a so-called
group grading. The most notorious case of a group grading is the
$\mathbb Z_2$-grading introduced by E.~In\"{o}n\"{u} and E.~Wigner
when decomposing a Lie algebra $L$ into two non-zero grading
subspaces $L_0$ and $L_1$, where
\begin{equation}\label{gradaceNaDva}[L_0, L_0]\subseteq L_0,
\qquad [L_0, L_1]\subseteq L_1, \qquad [L_1, L_1]\subseteq
L_0.
\end{equation}

A grading $\Gamma : L= \oplus_{j\in\mathcal J} L_j$ is called a {\bf
group grading} if the index set $\mathcal J$ can be embedded into  a
semigroup $G$ (whose binary operation is denoted by $+$), and, for
any pair of indices $j,k\in\mathcal J$, it holds that
\begin{equation}\label{group-grading-commutation}
[L_j, L_k] \subseteq L_{j+k}.
\end{equation}
Since we allow even trivial subspaces in decomposition of $L$, as
index set of the group grading may be used directly the semigroup
$G$. In this case we will speak about $G$-grading $\Gamma$.

\bigskip

A grading $\Gamma: L = \oplus_{i\in \mathcal J} L_i$ of a Lie
algebra $L$ is a starting point for study of {\bf graded
contractions} of the Lie algebra. This method for finding
contractions of Lie algebras was introduced in \cite{MoPa, MuPa}.
Since we will focus in this paper on group gradings only, we
assume  in the sequel  that the indices of grading subspaces
belong to a group $G$, i.e., $\Gamma$ is a $G$-grading of $L$.

In this type of contraction, we define new Lie brackets by
prescription

\begin{equation}\label{zavorky}[ x,y]_{new} := \varepsilon_{j,k}[x,y], \ \hbox{where}\ x\in
L_j, y\in L_k.
\end{equation}

The complex or real parameters $\varepsilon_{j,k}$, for $j,k\in
G$, must be determined in such a way that the vector space $L$
with the binary operation $[.,.]_{new}$ forms again a Lie algebra.
Antisymmetry of Lie brackets  demands that $\varepsilon_{j,k} =
\varepsilon_{k,j}$. If moreover, the coefficients
$\varepsilon_{j,k}$  fulfill a
 system of quadratic equations:
\begin{equation}\label{epsilony}
\varepsilon_{i,j}\varepsilon_{i+j,k}
=\varepsilon_{j,k}\varepsilon_{j+k,i}=\varepsilon_{k,i}\varepsilon_{k+i,j}\quad
\hbox{for all} \ \ i,j,k \in  G
\end{equation}
then the vector space  $L$ with  new  brackets $[ x,y]_{new}$
satisfies the Jacobi identities as well. This new Lie algebra will
be denoted   by $L^\varepsilon$.

\begin{ex}\label{priklad1} For a  $\mathbb{Z}_2$-grading of a Lie algebra $L$, the
system of equations \eqref{epsilony} has a very simple form
$$ (\varepsilon_{00}-\varepsilon_{01})\varepsilon_{01} = 0=
(\varepsilon_{00}-\varepsilon_{01})\varepsilon_{11}
$$
There exist infinitely many solutions  $\varepsilon =
(\varepsilon_{jk})$ of this system. Nevertheless for many
solutions, the contracted algebras $L^\varepsilon$ are isomorphic.
It can by shown  that only four solutions
$$
(\varepsilon_{jk}) =
 \left( \begin{array}{cc}1 &1\\1 & 0 \end{array}  \right)\,,  \quad
    \left( \begin{array}{cc}1 &0\\0 & 0\end{array}  \right)\,,
    \quad
  \left( \begin{array}{cc}0 &0\\0 & 1 \end{array}  \right)\,,
  \quad{\rm and} \quad
 \left( \begin{array}{cc}0 &0\\0 & 0\end{array}  \right)
$$
 give mutually non-isomorphic Lie algebras $L^\varepsilon$.
The contracted  algebra obtained by the first solution is the
semidirect sum of $L_0$ with a commutative algebra $L_1$. The
second solution is the direct sum of $L_0$ and the commutative
algebra $L_1$. The third solution leads to the Lie algebra of the
Heisenberg type, or equivalently, it is a central extension of the
commutative algebra $L_1$ by the commutative algebra $L_0$. The
fourth solution is an Abelian Lie algebra.

\end{ex}

\subsection{Representations of graded contractions}

Let us focus on the question of representation of the contracted
Lie algebra $L^\varepsilon$. We will describe a method
\cite{PaTo}, which enables to find a representation of
$L^\varepsilon$ by modifying a given representation of the
original algebra $L$.

\begin{de}\label{compatibilitaD}Let $r: L \mapsto {\rm End}\,V$ be a
representation of the Lie algebra $L$  and $\Gamma: L =
\oplus_{i\in G} L_i$ be its $G$-grading. We say that {\bf the
representation }$r$ is {\bf compatible with the $G$-grading}, if
there exists a decomposition of the vector space $V$ into a direct
sum $ V = \oplus_{i\in \mathcal J} V_i$ such that
\begin{equation}\label{compatibilita}
r(X_i)V_j \subset V_{i+j} \qquad \hbox{for each} \  \ i,j \in G
 \ \ \hbox {and any} \ \ X_i\in L_i \,.
\end{equation}
\end{de}

Suppose we are given a representation $r$ of $L$ compatible with
$G$-grading. We are looking for a representation $r^\varepsilon$
of a contracted Lie algebra $L^\varepsilon$. Let us define
\begin{equation}\label{reprezentace}
r^\varepsilon(X_i)v_j:= \psi_{i,j} \,r(X_i)v_ j \qquad \hbox{for
each} \  \ i,j \in G\,,
 \ \ \hbox { any} \ \ X_i\in L_i  \hbox { and any} \ \ v_j\in V_j,
\end{equation}
where $ \psi_{i,j}$ are unknown parameters. The requirement that
$r^\varepsilon$ is a representation of $L^\varepsilon$  formally
means
$$  r^\varepsilon\bigl([X_i, X_j]_{new}\bigr)v_k =
[r^\varepsilon(X_i), r^\varepsilon(X_j)]v_k =\bigl(r^\varepsilon(X_i) r^\varepsilon(X_j)- r^\varepsilon(X_j),
r^\varepsilon(X_i)\bigr)v_k
$$
for any $X_i \in L_i, X_j\in L_j$, and $v_k\in V_k$. Using equations
\eqref{zavorky}, and \eqref{reprezentace} and the relation
\eqref{compatibilita} we obtain
$$
\psi_{j,k}\psi_{i,j+k} r(X_i)r(X_j) - \psi_{i,k}\psi_{j,i+k}
r(X_j)r(X_i) =\varepsilon_{i,j}\psi_{i+j,k}r([X_i,X_j])
$$
Since $r$ is a representation of $L$, we know that $ r(X_i)r(X_j) -
r(X_j)r(X_i)=r([X_i,X_j])$. Therefore,  the choice of parameters
$\psi_{i,j}$ satisfying
\begin{equation}\label{RovniceProPsi}
\psi_{j,k}\psi_{i,j+k} = \psi_{i,k}\psi_{j,i+k}
=\varepsilon_{i,j}\psi_{i+j,k}
\end{equation}
implies that $r^\varepsilon$  defined by \eqref{reprezentace} is
representation of the contracted Lie algebra $L^\varepsilon$.
Comparing \eqref{RovniceProPsi} and \eqref{epsilony} we see that
the systems of quadratic equations for parameters $\psi_{i,j}$ has
at least one solution, namely  $\psi_{i,j} = \varepsilon_{i,j}$
for each pair $i,j$. Therefore the mapping  $r^\varepsilon:
L^\varepsilon \mapsto {\rm End}\,V$ defined by
\eqref{reprezentace} is a representation of the graded Lie algebra
$L^\varepsilon$. Usually, there exist also other  solutions of the
system \eqref{RovniceProPsi}, and therefore more representations
of the same contracted algebra $L^\varepsilon$.
\begin{ex}\label{priklad2}
Consider a  $\mathbb{Z}_2$-grading of a Lie algebra $L$ and its
 representation $r$ which is
compatible with the grading.  For the corresponding decomposition
of the vector space  $V = V_0\oplus V_1$ we may construct a basis
$\mathcal{B}$ of $V$ by putting together the bases of $V_0$ and
$V_1$. In such a basis $\mathcal{B}$, the grading relations
 \eqref{gradaceNaDva} give
explicitly
$$ r(X_0) =\left( \begin{array}{cc}A(X_0) &0\\0 & B(X_0)
 \end{array}  \right)\qquad \hbox{and}\qquad  r(X_1) =\left( \begin{array}{cc}0&C(X_1)
 \\D(X_1)&0
 \end{array}  \right)\,.$$
In the sequel, we will illustrate all notions on the Lie algebra
$L^\varepsilon$, which is obtained by contraction  from
$\mathbb{Z}_2$-grading of a Lie algebra $L$ by the first solution
$$
(\varepsilon_{jk}) = \left( \begin{array}{cc}1 &1\\1 & 0
 \end{array}  \right)$$
given in Example \ref{priklad1}. For this Lie algebra
$L^\varepsilon$ the commutation relations have the form

\medskip

\centerline{$ [x,y]_{new} = [x,y]$, if $x,y\in L_0$ or if  $x\in
L_0, y\in L_1$ \ \ and \ \ $ [x,y]_{new} = 0$, if $ x, y\in L_1$.}

\medskip

\noindent  In this case the system of equations
\eqref{RovniceProPsi} is
$$ \psi_{00}\psi_{00} = \psi_{00}, \quad
\psi_{10}\psi_{01}=\psi_{00}\psi_{10} = \psi_{10}\,,$$
$$ \psi_{01}\psi_{01} = \psi_{01}, \quad
\psi_{11}\psi_{00}=\psi_{01}\psi_{11} = \psi_{11}\,,$$
$$ \psi_{10}\psi_{11} = 0\,.$$
All solutions (up to equivalence of representations) of this
system are
$$
(\psi_{jk}) = \left( \begin{array}{cc}1 &1\\1 & 0
 \end{array}  \right)\,,  \quad
 \left( \begin{array}{cc}1 &1\\0 &1 \end{array}  \right)\,, \quad
  \left( \begin{array}{cc}1 &1\\0 &0\end{array}  \right)\,,  \quad
   \left( \begin{array}{cc}1 &0\\0 & 0\end{array}  \right)\,, \quad
  \left( \begin{array}{cc}0 &1\\0 & 0 \end{array}  \right)\,, \quad  {\rm and} \quad
  \left( \begin{array}{cc}0 &0\\0 & 0\end{array}  \right)
$$
The representations $r^\varepsilon$ of the contracted Lie algebra
$L^\varepsilon$ in the chosen basis $\mathcal{B}$ of the vector
space $V$ have the form
$$ r(X_0) =\left( \begin{array}{cc}\psi_{00}A(X_0) &0\\0 & \psi_{01}B(X_0)
 \end{array}  \right)\qquad \hbox{and}\qquad  r(X_1) =\left( \begin{array}{cc}0&\psi_{11}C(X_1)
 \\ \psi_{10}D(X_1)&0
 \end{array}  \right)\,,$$
where for parameters $(\psi_{ij})$ one may choose one of the six
solutions. Let us mention that only two first solutions are
interesting since the elements of subalgebra $L_1$ are in the
remaining solutions represented by zero operators.
\end{ex}

\subsection{Some open problems}

 The four notions introduced above: grading, group grading,
graded contraction and representation compatible with G-grading,
immediately lead to many questions.

\begin{enumerate}
\item\label{otazka1} How many inequivalent gradings has a given
Lie algebra $L$?

\item\label{otazka2} Do notions of grading and group grading
coincide? On which Lie algebras?

\item\label{otazka3} Is the semigroup $G$ assigned to a group
grading $\Gamma$ uniquely? If not, does there exist a semigroup
$G$ to be preferred?

\item\label{otazka4} Does there exist a $G$-grading of $L$ for a
given Lie algebra $L$ and a given semigroup $G$? How many
inequivalent $G$-gradings of $L$ one can find?

\item How many non-isomorphic graded contractions can be found for
a given $G$-grading of $L$?

\item Which irreducible  representations of a Lie algebra $L$ are compatible with
its $G$-grading?

\end{enumerate}

\noindent No question of this list has found a satisfactory answer
yet. Let us briefly summarize the achievements of numerous papers
in this direction.

\medskip

 It seems, from the theoretical point of view, that the most
important question to be solved is the second question on our
list.  In \cite{Elduque}, Elduque contradicted the longstanding
belief that any grading is a group grading. He found a Lie algebra
of dimension 17 and described its grading which cannot be indexed
by any semigroup $G$. The algebra which served to Elduque as
counterexample is not simple. But even in case of simple
finite-dimensional Lie algebras the situation is not clear. Up to
now no non-group grading of a simple Lie algebra was found.
Non-existence of non-group grading for any simple Lie algebra was
claimed in the seminal work \cite{PaZa}. But Elduque's result
showed a gap in the proof of this statement. Therefore, the lists
of fine inequivalent gradings of simple Lie algebras which were
given in \cite{HaPaPe}, \cite{D4}, \cite{G2} and \cite{F4} are
complete only as lists of group gradings. To show that these lists
are complete in general sense, one needs to prove
\medskip

\centerline{{\bf  Conjecture:} \ {\it Any grading of a simple Lie
algebra is a group grading.}}

\medskip

It is not difficult to answer negatively the first part of the
third  question. Namely,  for any group grading $\Gamma$ there
exist more non-isomorphic semigroups suitable for labeling
subspaces $L_i$. As shown in \cite{G2}, any group grading of a
simple finite dimensional Lie algebra  can be indexed by a
finitely generated Abelian group $G_I$. Moreover,  surjective
homomorphisms of $G_I$ are in one-to-one correspondence with
coarsenings of the grading $\Gamma$. The Abelian group $G_I$ with
this  property is given uniquely up to isomorphisms.

From the point of view of the result Draper and Mart\'{\i}n in
\cite{G2}, the question \ref{otazka4} seems to be less difficult
for simple finite-dimensional Lie algebras. We can restrict our
considerations to Abelian finitely generated groups, i.e. \  to
additive groups $G=\mathbb{Z}_{n_1}\otimes\,  \mathbb{Z}_{n_2}
\otimes\ldots \otimes\, \mathbb{Z}_{n_r} \otimes \,\mathbb{Z}^k$ \
for some integers $n_1,n_2, \ldots n_r, k$. Nevertheless, till now
the answer is known only for the simplest groups $G$, namely
$\mathbb{Z}_n$, $\mathbb{Z}$ and $\mathbb{Z}_2 \otimes
\mathbb{Z}_2$,  \cite{Bahturin}.

The task to decide whether two  Lie algebras given by their
structural constants are isomorphic or not is far from being
simple. Therefore only several examples of low-dimensional
algebras and their graded contractions are well understood,
\cite{HrNoTo}.

As we have already said,  the first steps in construction of
representation of graded contractions were made in \cite{MuPa},
\cite{MoPa} and \cite{PaTo}. In these works,   authors showed for
specific type of $G$-gradings  of simple  Lie algebras that any
irreducible representation is compatible with  $G$-grading.
Moreover, they gave also a recipe how to find suitable
decomposition of vector space $V$ satisfying
\eqref{compatibilita}.

\medskip

The article is organized as follows. For representation theory the
distinction of gradings and group gradings is important. This open
problem is approached in Section  \ref{naDva}. It contains a
modest improvement of the just mentioned conjecture: we show that
the conjecture is true if a grading consists of two subspaces
only.

The central part of the paper (Sections 3 and 4) is devoted to
representations compatible with a grading. Explicit results are
obtained for finite-dimensional representations of
$sl(3,\mathbb{C})$ compatible with $\mathbb{Z}_2$-gradings
generated either by an inner automorphism of order 2 or by an
outer automorphism of order 2.

Our goal is to enlarge the family of gradings of L for which one
can decide about compatibility with the given representation of L.
These results are illustrated on the simple Lie algebra
$sl(3,\mathbb{C})$.

\section{Grading versus group grading}\label{naDva}

In this section we concentrate on simple Lie algebras of finite
dimension, say $k\in \mathbb{N}$. In particular, we will use the
fact that $L$ is perfect, i.e.
 \begin{equation}\label{perfektni} [L,L]
:= \{ [x,y]\, \mid \,x,y \in L\} = L\,.
 \end{equation}
At first we recall several ingredients needed in the proof of the
theorem. We will work with the adjoint representation which  to
any $ x\in L$  assigns the linear operator $M_x$ given on the
vector space $L$  by the prescription
 \begin{equation}\label{adjungovana}
 M_x \, y:= [x,y]\,.
 \end{equation}
Since $L$ is a simple algebra, the adjoint representation  is
irreducible and according to the Burnside theorem
\begin{equation}\label{vsechno}
 {\rm dim}\,\{ M_{x_1}M_{x_2}\ldots M_{x_n} \,\mid \ n\in \mathbb{N}, \ x_1,x_2,\ldots, x_n \in L \}_{\rm lin} =
 k^k
 \end{equation}

The following theorem is a special case of the conjecture we
mentioned in Section \ref{zacatek}.

\begin{thm}\label{musigrupa} Let $L$ be a simple finite dimensional Lie algebra and
let
$$
\Gamma: \   L= L_a \oplus L_b $$ be its grading into two nontrivial
subspaces. Then $\Gamma$ is a $\mathbb{Z}_2$-grading.

\end{thm}
\bp  Since $\Gamma$ is a grading, there exist letters $x,y,z\in
\{a,b\}$ such that

 \begin{equation}\label{moznosti}
 [L_a,L_a]\subset L_x, \quad   [L_a,L_b]\subset L_y, \quad {\rm and}
 \quad  [L_b,L_b]\subset L_z \,.
 \end{equation}
The property \eqref{perfektni} guarantees that both  letters $a$
and $b$ occur among $x,y,z$, i.e.,
 \begin{equation}\label{pokryti}
\{a,b\} = \{x,y,z\}
 \end{equation}
According to
 occurrences of   inclusions  $[L_c,L_c]\subset L_c$,  we will discuss   separately three
 cases:

 \medskip

$\bullet$ There exists only  one letter $c \in \{a,b\}$,  such
that $[L_c,L_c]\subset L_c$. Without loss of generality, we may
assume $c=a$. Then necessarily for the second index
$[L_b,L_b]\subset L_a$. Using \eqref{pokryti}, the relations
\eqref{moznosti} have the form
$$
[L_a,L_a]\subset L_a, \quad   [L_a,L_b]\subset L_b, \quad {\rm and}
 \quad  [L_b,L_b]\subset L_a
$$
(which is equivalent to $[L_a,L_a]\subset L_a, \, [L_a,L_b]\subset
L_b,$ and $[L_b,L_b]\subset L_a $).

If we identify $a$ with $0$ and $b$ with $1$ in the group
$\mathbb{Z}_2$, we see, that our grading fulfills $[L_i,L_j]
\subset L_{i+j}$, as desired.

$\bullet$ For both letters $c=a$ and $c=b$ we have $[L_c,L_c]\subset
L_c$. Since the role of $a$, and $b$ is symmetric, we may write
 \begin{equation}\label{moznosti2}
[L_a,L_a]\subset L_a, \quad   [L_a,L_b]\subset L_a, \quad {\rm and}
 \quad  [L_b,L_b]\subset L_b
 \end{equation}
Let us denote ${\rm dim}\, L_a = k_a$ and ${\rm dim}\, L_b = k_b$.
Clearly $k_a+k_b=k$.  Since $ L= L_a \oplus L_b $,  a basis of $L_a$
and a basis of $L_b$ form together a basis of $L$. Let us denote
this basis $\mathcal{X}$. Let us consider the form matrices  of
operators $M_x$ in this base. In fact we will identify $M_x$ with
its  matrix. The  definition \eqref{adjungovana} and the relations
\eqref{moznosti2} imply
$$
M_x = \left( \begin{array}{cc} M^{11}_x & M^{12}_x \\0 & 0
 \end{array}  \right)  \quad\hbox{for } \ x \in L_a \ \ \qquad {\rm and}\qquad \ \ M_y = \left( \begin{array}{cc} M^{11}_y & 0\\0 &
 M^{22}_y
 \end{array}  \right) \quad \hbox{for } \ y \in L_b
$$
It means that $M_x$ and $M_y$ are both block upper triangular
 matrices. This contradicts
\eqref{vsechno}.

$\bullet$ It remains to discuss the case
 \begin{equation}\label{moznosti3}
[L_a,L_a]\subset L_b, \quad   [L_a,L_b]\subset L_b, \quad {\rm and}
 \quad  [L_b,L_b]\subset L_a\,.
 \end{equation}
  Now $M_x$
and $M_y$ have the form
 \begin{equation}\label{tvar}
M_x = \left( \begin{array}{cc}0 & 0\\ M^{21}_x & M^{22}_x
 \end{array}  \right)  \quad\hbox{for } \ x \in L_a \ \ \qquad {\rm and}\qquad \ \
 M_y = \left( \begin{array}{cc}  0& M^{12}_y\\
 M^{21}_y& 0
 \end{array}  \right) \quad \hbox{for } \ y \in L_b
 \end{equation}
 We may assume  $[L_a,L_a]\neq 0$, otherwise we may write $[L_a,L_a]
= 0 \subset L_a $ and this case was already discussed. It means
that there exists $x_0\in L_a$ such that  $M^{21}_{x_0}\neq 0$.
Denote the rank of $M^{21}_{x_0}$ by $h\geq 1$. The bases of $L_a$
and $L_b$ were chosen arbitrarily. Now we may assume without loss
of generality that we have chosen these bases in such a way that
$$M^{21}_{x_0} = \left( \begin{array}{cc}I_h & 0\\ 0 & 0
 \end{array}  \right),$$
where $I_h$ is the unit matrix of size $h\times h$, and the blocks
of zeros complete the matrix to the size of $M^{21}_{x_0}$ which
is $k_b\times k_a$.

Let us compute for any $y\in L_b$ the commutator  $[M_{x_0}, M_y]$.
We obtain
$$  \left( \begin{array}{cc}0 & 0\\ M^{21}_{x_0} & M^{22}_{x_0}
 \end{array}  \right)\left( \begin{array}{cc}  0& M^{12}_y\\
 M^{21}_y& 0
 \end{array}  \right) - \left( \begin{array}{cc}  0& M^{12}_y\\
 M^{21}_y& 0
 \end{array}  \right)\left( \begin{array}{cc}0 & 0\\ M^{21}_{x_0} & M^{22}_{x_0}
 \end{array}  \right) = \left( \begin{array}{cc}- M_y^{12}M_{x_0}^{21}&-M^{12}_y M^{22}_{x_0}\\
  M^{22}_{x_0} M^{21}_y & M_{x_0}^{21}M_y^{12}
 \end{array}  \right)
$$
Since $[x_0,y]\in L_b$, we have $$M_{x_0}^{21}M_y^{12} = 0 =  \left(
\begin{array}{cc}I_h & 0\\ 0 & 0
 \end{array}  \right) M_y^{12}$$
It implies that  for any $y\in L_b$ the first $h$ rows of the
matrix $M_y^{12}$ and therefore also of the matrix $M_y$ contain
only zeros. Taking into account  the form of $M_x$, we see that
any element of the Lie algebra is represented by the matrix whose
first row is zero and this contradicts \eqref{vsechno}.
 \ep

\section{Representations compatible with grading}
\subsection{Group gradings and automorphisms}

The  simplest way how to find a group grading of a Lie algebra is
to decompose the vector space $L$ into eigensubspaces of a
diagonalizable automorphism $g\in \mathcal{A}ut\, L$,
\cite{HaPaPe}. For any pair of its eigenvectors $x_\lambda$ and
$x_\mu$ corresponding to eigenvalues $\lambda$ and $\mu$,
respectively, we have
$$g([x_\lambda, x_\mu]) = [g(x_\lambda),g( x_\mu)] = \lambda \mu [x_\lambda,
x_\mu])\,.$$ Thus the commutator $[x_\lambda, x_\mu]$ is either
zero or an eigenvector corresponding to the eigenvalue $\lambda
\mu$. Let us denote by $\sigma(g)$ the spectrum of the
automorphism $g$ and by $L_\lambda$ the eigensubspace
corresponding to $\lambda \in \sigma(g)$. The decomposition
 \begin{equation}\label{jeden}
 \Gamma: \ L =\ \bigoplus_{\lambda \in
 \sigma(g)}L_\lambda
 \end{equation}
 is a group grading, where as a
semigroup $G$ one can use the multiplicative semigroup generated
by the spectrum of $g$. Clearly, if $h\in \mathcal{A}ut\, L$ then
the decomposition of $L$ into eigensubspaces of the automorphism
$hgh^{-1}$ is $ L =\ \bigoplus_{\lambda \in
 \sigma(g)}h(L_\lambda)$, i.e., the grading given by automorphisms $g$ and
 $hgh^{-1}$ are equivalent. Therefore, the automorphisms $g$ and
 $hgh^{-1}$ are called equivalent as well.  Note however that  different inequivalent
 automorphisms may
give the same grading.

Similarly, if $g_1$, $g_2$, \ldots, $g_r$ are mutually commuting
automorphisms of $L$, then the decomposition of $L$ into common
eigensubspaces of all these automorphisms is a group grading of
$L$. The suitable semigroup for indexing of this grading is $G_1
\otimes G_2 \otimes \ldots \otimes G_r$, where  $G_i$ is the
semigroups generated by the spectrum of $g_i$.

Furthermore, for Lie algebras over the complex field $\mathbb{C}$,
any group grading can be obtained by the described procedure.  Let
us stress, that this is not the case for real Lie algebras.

\subsection{Group grading determined by one automorphism}
Let $\Gamma$ be a grading of the form \eqref{jeden}, i.e. obtained
by decomposition of $L$ into eigensubspaces of a single
automorphism $g$. We may assume that $g$ has finite order, say
$g^k=Id$. For its spectrum we have
\begin{equation}\label{grupa}\sigma(g)\subset \{ e^{i\frac{2\pi}{k}\ell} \mid
\ell = 0,1,2,\ldots, k-1\} =: G.
 \end{equation}
 It means that
$\Gamma$ is a $G$-grading.  Let us consider an irreducible
$d$-dimensional representation  $r$ of the Lie algebra $L$. Our
aim is to discuss the question of compatibility of $r$ with the
$G$-grading.

Let $R_g$ be a non-singular  matrix in $\mathbb{C}^{d\times d}$
such that
\begin{equation}\label{simuluje1}
r(g(x)) = R_gr(x)R_g^{-1}\quad {\rm for\  all} \ \ x\in L\,.
\end{equation}
 As $g^k = Id$, the previous equality gives
$$r(x) = r(g^k(x)) = R_g^kr(x)R_g^{-k}\, \quad {\rm or } \quad [R_g^k,r(x)] = 0\quad \ {\rm for\  all} \ x \in L\,.$$
Since the representation  $r$ is irreducible, by Schur's lemma $
R_g^k = \alpha \,Id$, for some $\alpha \in \mathbb{C}$.  Of
course, any nonzero multiple of $R_g$ satisfies the relation
\eqref{simuluje1} as well. Therefore without loss of generality,
we may assume that
\begin{equation}\label{simuluje2}
R_g^k = Id\,, \quad \hbox{where $k$ is the order of the
automorphism $g$.}
\end{equation}
This normalization guarantees that the spectrum of the matrix
$R_g$ and the spectrum of the automorphism $g$  belong to the same
group $G$. In particular, since $R_g^k$ is the identity, the
matrix $R_g$ is diagonalizable. Let  $ V = \oplus_{\lambda \in G}
V_\lambda$ denote  the decomposition of the column space
$\mathbb{C}^d$ into eigensubspaces of the matrix $R_g$. We will
show, that this decomposition is exactly the decomposition
required in the definition \ref{compatibilitaD}.

Let us consider some $\mu \in \sigma(g)$ so that  $g(x_\mu) = \mu
x_\mu$ for all $x_\mu \in L_\mu$. From \eqref{simuluje1} for
$x=x_\mu$, one can deduce
$$ r(g(x_\mu))R_g = r(\mu x_\mu)R_g =  \mu \,r( x_\mu)R_g= R_gr(x_\mu)$$
and after multiplying by the column vector $v_\lambda \in
V_\lambda $
$$ \mu\, \lambda\, r( x_\mu)v_\lambda  = R_gr(x_\mu) v_\lambda\,.$$
The last equality means that the column  $r( x_\mu)v_\lambda$ is
either zero, or it is  an eigenvector of the matrix $R_g$
corresponding to the eigenvalue $\mu\,\lambda$. Therefore
$$r( x_\mu)V_\lambda \subset   V_{\mu\lambda}
\quad \hbox{ for any $\lambda , \mu \in G$ \ and\  any  $x_\mu\in
L\mu$ }\,.$$ This is the relation \eqref{compatibilita}, just
written in the multiplicative form. Of course, our multiplicative
group $G$ defined in \eqref{grupa} is isomorphic with the additive
group $\mathbb{Z}_k$.

We have seen, that the matrix $R_g$ with the properties
\eqref{simuluje1} and \eqref{simuluje2} guarantees the
compatibility  of the grading of $L$  with the representation of
the Lie algebra $L$. Such matrix $R_g$ will be called {\bf
simulation matrix } of the automorphism $g$. The matrix $R_g$
depends on the chosen automorphism $g$ and on the chosen
representation $r$. The idea how to find the simulation matrix is
more straightforward if $g\in \mathcal{A}ut\, L$ is an inner
automorphism. In this case it is natural to search for $R_g$ among
matrices in the representation of the corresponding Lie group.
This idea was already presented  in \cite{MoPa} and \cite{PaTo},
where $R_g$ was a representation of power of an element of finite
order \cite{Kac}. Nevertheless, we show that it is possible to
find the simulation matrix $R_g$ even for an outer automorphism
$g$ as well. In the sequel, we will concentrate on the Lie
algebras $sl(n, \mathbb{C})$. The reason is, that these algebras
(with the exception of $o(8, \mathbb{C})$) are the only simple
classical Lie algebras for which the group of automorphisms
contains  an outer automorphism as well.

\section{Representations of $sl(n, \mathbb{C})$ compatible with $\mathbb{Z}_2$-grading}
According to Theorem \ref{musigrupa} any graded decomposition  of
$sl(n, \mathbb{C})$ into two parts is a $\mathbb{Z}_2$-grading.
Any such grading is uniquely related by an automorphism of order
$2$. We will identify the Lie algebra $sl(n, \mathbb{C})$ with $\{
X\in \mathbb{C}^{n\times n}\mid \,{\rm tr} X=0\}$.

Let us first recall the structure of ${\mathcal A}ut\, sl(n,
\mathbb{C})$ as described in \cite{automorfizmy}:

\begin{itemize}
\item for any inner automorphism $g$ there exists a matrix $A\in
SL(n, \mathbb{C}): = \{ A\in \mathbb{C}^{n\times n}\mid \det
A=1\}$ such that
$$ g(X) = Ad_A X = AXA^{-1} \quad \hbox{for any} \ X \in sl(n,
\mathbb{C})\,;$$ \item the mapping given by the prescription
$$Out_I X:= -X^T\quad \hbox{for any} \ X \in sl(n,
\mathbb{C})\,$$ is an outer automorphism of order 2;
 \item any outer automorphism $g$ is a composition of  an inner automorphism and the
 automorphism $Out_I$.
\end{itemize}

The second ingredient for construction of simulating matrices of
automorphisms is the knowledge of  finitedimensional irreducible
representations of $sl(n, \mathbb{C})$. These representations are
well described by Gel'fand-Tseitlin formalism \cite{Tsetlin, LePa,
Barut}. Any irreducible representation $r$ of $sl(n, \mathbb{C})$
is in one-to-one correspondence  with
 an $n$-tuple $(m_{1,n}, m_{2,n}, \ldots , m_{n,n})$
of non-negative integer parameters $m_{1n}\geq m_{2n}\geq \ldots
\geq   m_{nn}=0$. The dimension of the representation space of $r=
r(m_{1,n}, m_{2,n}, \ldots , m_{n,n})$ is given by the number of
triangular patterns
$$ {\bf m}=\left( \begin{array}{ccccccccc}m_{1,n}& &m_{2,n}&& m_{3,n}&&\ldots &&
m_{n,n}\\
&m_{1,n-1} &&m_{2, n-1}&&m_{3,n-1} &\ldots&
m_{n-1,n-1}&\\
&&m_{1,n-2} &&m_{2, n-2}& \ldots&
m_{n-2,n-2}&&\\
&&&\vdots  && \vdots&
\vdots &&\\
& &&m_{1,2}&  &m_{2,2}& &&
\\
& &&&m_{1,1}  && &&
\\
 \end{array}  \right)
$$
in which the numbers  $m_{i,j} \in \mathbb{Z}$ satisfy
$m_{i,j+1}\geq m_{i,j}\geq m_{i+1, j+1}$ for all $1\leq i\leq
j\leq n-1$. To any such pattern ${\bf m}$,  we assign the basis
vector $\xi({\bf m})$. The representation $r$ is fully determined
by the action $r(E_{k\ell})$ on all basis vectors $\xi({\bf m})$
for any $k,\ell = 1,2,\ldots,n$. (We have adopted notation
$E_{k\ell}$ for the matrices $n\times n$ with elements
$\bigl(E_{k\ell}\bigr)_{ij} = \delta_{ik}\delta_{\ell j}$.) This
action  can be found e.g. in \cite{Tsetlin}. For reader
convenience  the representation is described in Appendix.

\bigskip

\subsection{Inner automorphisms of order two}

Any inner automorphism $g$ of order two is associated by equality
$g = Ad_A$ with a group element $A \in SL(n, \mathbb{C})$ such
that $A$ does not belong to the center ${\mathcal{Z}}[SL(n,
\mathbb{C})]$ and $A^2$ belongs to the center. If we denote
$\omega = e^{\frac{i\pi}{n}}$, then the center can be written
explicitly $\mathcal{Z}[Sl(n, \mathbb{C})]= \{ \omega^{2\ell} I_n
\mid \ell = 0,1,\ldots, n-1\}$. A simple calculation shows that
any such element  $A \in SL(n, \mathbb{C})$  is  up to equivalence
one of the matrices
 $$A_{n,s} := \omega^{\eta(s)} \left( \begin{array}{cc}I_{n-s} & 0\\ 0 &
 -I_s
 \end{array}  \right)\quad  \hbox{where } \ s=0,1,\ldots, \lfloor\tfrac{n}{2}\rfloor
  \ \ \hbox{and}\ \  \eta(s) =
  \left\{ \begin{array}{ll}0 & \hbox{if $s$ is even}\\ 1 & \hbox{if $s$ is
  odd}
 \end{array}   \right.$$
These matrices may be rewritten by using elements of the Lie
algebra $sl(n, \mathbb{C})$    as follows:
$$A_{n,s} = \exp(X_{n,s}) \ \ \hbox{ with}  \ \ X_{n,s} =
i\pi \left( \begin{array}{cc}\frac{\eta(s)}{n}\,I_{n-s} & 0\\ 0 &
\frac{\eta(s)}{n}\,I_s + M_s \
 \end{array}  \right)\,,$$
where $ M_s= diag (-1, 1, -1, \ldots, (-1)^s) \in
\mathbb{C}^{s\times s}$.  One can  use notation of $E_{kk}$ and
write
\begin{equation}\label{strucne}
X_{n,s}= i\pi \left( \frac{\eta(s)}{n}\sum_{k=1}^{n} E_{kk} +
\sum_{k=n-s+1}^{n} (-1)^{n-s+1-k}E_{kk} \right)\,.
\end{equation}
If $r$ is any representation of the Lie algebra  $sl(n,
\mathbb{C})$, then $R_{A_{n,s}}:=\exp(r(X_{n,s}))$ satisfies
$$ R_{A_{n,s}}r(X)\bigl(R_{A_{n,s}}\bigr)^{-1} =
r(A_{n,s}XA_{n,s}^{-1}) = r(Ad_{A_{n,s}}X)\qquad {\rm and}\qquad
\bigl( R_{A_{n,s}}\bigr)^2=Id\,.$$ Therefore,
 the matrix $R_{A_{n,s}}:=\exp(r(X_{n,s}))$ is  the simulation
 matrix of the inner automorphism $g= Ad_{A_{n,s}}$.
We have shown
\begin{thm} \label{vnitrniJde} Any  $\mathbb{Z}_2$-grading of the Lie algebra
$sl(n, \mathbb{C})$ obtained by an  inner automorphism and any
irreducible  representation  of  $sl(n, \mathbb{C})$  are
compatible.
\end{thm}

Using \eqref{strucne} and the explicit  form of the
Gel'fand-Tseitlin representation we obtain for any basis vector
$\xi({\bf m})$

 $$r(X_{n,s})\xi({\bf m}) = i\pi \left( \frac{\eta(s)}{n} r_n({\bf m})+
 2\sum_{k=1}^{s-1}(-1)^{k-1} r_{n-s+k}({\bf m}) - r_{n-s}({\bf m}) - (-1)^{\eta(s)}r_{n}({\bf
 m})
 \right)\xi({\bf
 m})\,.
$$
So we arrived at the  explicit form of the simulation matrix of
the automorphism $g= Ad_{A_{n,s}}$
$$R_{A_{n,s}}\xi({\bf
 m}) = e^{i\pi\left( \bigl(\tfrac{\eta(s)}{n} -1\bigr)r_{n}({\bf
 m})- r_{n-s} ({\bf
 m})\right)}\xi({\bf
 m})
$$

\subsection{Outer automorphisms of order two}
Any outer automorphism of order two on $sl(n, \mathbb{C})$ is up
to equivalence the automorphism  $Out_I(X) = -X^T$, and thus we
will focus only on it.

It is well known, that for an irreducible representation  $r$
characterized in the Gel'fand-Tseitlin formalism by the  $n$-tuple
$(m_{1,n}, m_{2,n}, \ldots , m_{n,n})$, the mapping $-r^T$ is an
irreducible representation as well. This representation is
equivalent to the so called contragradient representation  $r^c$
which is
 characterized by the $n$-tuple $(m_{1,n}', m_{2,n}',
\ldots , m_{n,n}')$, where
$$ m_{i,n}'= m_{1,n} - m_{ n-i+1,n} \quad \hbox{for} \ \  i=1,2,\ldots,n\,.$$
Let us consider  a triangular pattern ${\bf m}$ filled  by indices
$m_{i,j}$, $1\leq i\leq j\leq n$, and   associated with the basis
vector $\xi({\bf m})$ of  the representation $r$.  To any such
pattern ${\bf m}$, we may assign the unique triangular pattern
${\bf m}'$ with indices $m_{i,j}' :=m_{1,n} - m_{j-i+1,j}$. It is
easy to check that $m_{i,j}' $ satisfies the necessary
inequalities for ${\bf m}'$ to be a correct pattern of the
contragradient representation $r^c$. Let us define  the linear
mapping $J$ of the representation space of $r$ onto the
representation space of $r^c$ by
 $$ J\,\xi({\bf m}):= (-1)^{\sum_{i,j}m_{i,j}} \xi({\bf m}')\,.$$
Now from the formulae in the Appendix one sees that
 \begin{equation}\label{platiprotrnspozici}
 \bigl(r(E_{ij})\bigr)^T = r(E_{ji}) =
r(E_{ij}^{~T}) \,.
 \end{equation}
Using this fact one can prove by the  direct verification that the
mapping $J$ satisfies
 \begin{equation}\label{ekvivalence5}
 -J\,r^T(X) =  r^c(X) \,J\quad  \hbox{for any} \ \ X \in sl(n,
 \mathbb{C})\,.
 \end{equation}
Let us return to  our original task.  We are looking for the
simulation matrix of the  automorphism $g=Out_I$, i.e., we are
looking for a matrix $R_g$ of order two such that
$$r(Out_I(X)) = - r(X^T)= R_gr(X)R_g^{-1}\,.$$ According to
\eqref{platiprotrnspozici}, we have $r(X^T) = \bigl(r(X)\bigr)^T$
and  therefore the existence  of the  simulation matrix $R_g$
means equivalence of the representations $r$  and $-r^T$, i.e.
equivalence of $r$ and its contragradient  representation $r_c$.
The Gel'fand-Tseitlin result says that it is possible if and only
if $n$-tuples $(m_{1,n}, m_{2,n}, \ldots , m_{n,n})$ and
$(m_{1,n}', m_{2,n}', \ldots , m_{n,n}')$ coincide. In this case
the simulation matrix $R_g$ is equal to  $J$.  We have deduced
\begin{thm}\label{kdyjdevnejsi} A $\mathbb{Z}_2$-grading of the Lie algebra
$sl(n, \mathbb{C})$ obtained by an outer automorphism is
compatible with an irreducible  representation $r$  of  $sl(n,
\mathbb{C})$  if and only if the representation is self-
contragradient.
\end{thm}
If we do not insist on the irreducibility of the representation
$r$, the class of representations compatible with
$\mathbb{Z}_2$-grading  obtained by the automorphism $Out_I$ is
larger. Of course, if  for a representation $r_1$ it is possible
to find a simulation matrix $R^{(1)}$ and for  a representation
$r_2$  a simulation matrix $R^{(2)}$, then the direct sum
$R^{(1)}\oplus R^{(2)}$ is the simulation matrix for the direct
sum $r_1\oplus r_2$.  To avoid the discussion of all such obvious
cases, we will describe only those representations $r$ with
simulation matrices $R$, for which the operator set $\{R\} \cup \{
r(X)\mid X \in sl(n, \mathbb{C})\}$ is irreducible, whereas the
set $\{ r(X)\mid X \in sl(n, \mathbb{C})\}$ is reducible.

If $r_0$ is a $d$-dimensional irreducible representation of $
sl(n, \mathbb{C})$ then the $2d$-dimensional representation $r :=
r_0 \oplus \bigl(-r_0^{~T}\bigr)$ assigns to  $X$ the matrix
$$r(X) =
 \left( \begin{array}{cc} r_0(X) &0\\0 & -\bigl(r_0(X)\bigr)^T
 \end{array}  \right)
$$
and therefore
$$r(Out_I(X)) =
 \left( \begin{array}{cc} -\bigl(r_0(X)\bigr)^T &0\\0 & r_0(X)
 \end{array}  \right) =  \left( \begin{array}{cc}0 &I_d\\I_d & 0
 \end{array}  \right) r(X) \left( \begin{array}{cc}0 &I_d\\I_d & 0
 \end{array}  \right)\,.
$$
The matrix $\left( \begin{array}{cc}0 &I_d\\I_d & 0
 \end{array}  \right)$ is the simulation matrix of $Out_I$. It is
 easy to see that  the simulation matrix together with all $r(X)$
 form an irreducible set.

\subsection{$\mathbb{Z}_2$-grading of $sl(3, \mathbb{C})$}

Let us illustrate the conclusions of two previous sections on the
Lie algebra $sl(3, \mathbb{C})$. On this algebra there exists only
two inequivalent automorphisms of order two. In our notation
$g_1=Ad_{A_{31}}$ with
$$ A_{31} = \omega\left(
\begin{array}{ccc}1 &0&0\\0&1 & 0\\0&0&-1 \end{array}  \right)\,,
\quad \hbox{where}\quad \omega = e^{\frac{i\pi}{3}} $$ and
$g_2=Out_I$. The corresponding $\mathbb{Z}_2$-gradings are

$$ \Gamma_1: sl(3, \mathbb{C}) = \Bigl\{ \left( \begin{array}{ccc}
a&b&0\\c&d&0\\0&0&-a-d
 \end{array}  \right) \Big|\; a,b,c,d \in \mathbb{C}\Bigr\}\oplus
\Bigl\{ \left( \begin{array}{ccc} 0&0&a\\0&0&b\\c&d&0
 \end{array}  \right) \Big|\; a,b,c,d \in \mathbb{C}\Bigr\}\,,
$$

$$ \Gamma_2: sl(3, \mathbb{C}) = \Bigl\{ \left( \begin{array}{rrr}
0&a&b\\-a&0&c\\-b&-c&0
 \end{array}  \right) \Big|\; a,b,c \in \mathbb{C}\Bigr\}\oplus
\Bigl\{ \left( \begin{array}{ccc} a&b&c\\b&d&e\\c&e&-a-d
 \end{array}  \right) \Big|\; a,b,c,d,e \in \mathbb{C}\Bigr\}\,.
$$
The first grading $\Gamma_1$ is compatible  with any irreducible
representation. The simulation  matrix $R_{g_1}$ of the
automorphism $g_1=Ad_{A_{31}}$ acts on the Gel'fand-Tseitlin
triangular patterns as follows
$$
R_{g_1} \left(
\begin{gathered}
m_{13}\quad m_{23}\quad 0 \\
m_{12}\quad m_{22} \\
m_{11}
\end{gathered}\right)= e^{-\frac{2i\pi }{3}(m_{13}+m_{23})} e^{-i\pi
(m_{12}+m_{22})} \left(
\begin{gathered}
m_{13}\quad m_{23}\quad 0 \\
m_{12}\quad m_{22} \\
m_{11}
\end{gathered}\right)\,.
$$

The irreducible representations compatible with the second grading
are only  self-contragradient representations, i.e.,
representations $r=r(2\ell, \ell,0)$. In such representation, the
operator $J$ is defined by
$$ J\left(\begin{array}{ccc}
2\ell&\ell & \ 0\\
\multicolumn{3}{c}{\  m_{12}\ m_{22}}\\
& m_{11}
\end{array}
\right) =(-1)^{\ell+m_{12}+m_{22}+m_{11}} \left(\begin{array}{ccc}
2\ell&\ \ell & \quad 0\\
\multicolumn{3}{c}{\ \ 2\ell\!\!-\!\! m_{22}\ 2\ell\!\!-\!\! m_{12}}\\
& \ 2\ell \!\!-\!\! m_{11}
\end{array}
\right)\,.
$$

%
The lowest-dimensional non-trivial self-contragradient
representation is $r=r(2, 1,0)$. Its dimension is 8 and has the
following  explicit form on the basis vectors:
$$
R_{g_2}\left(
\begin{gathered}
2\ \ 1\ \ 0\\
2\ \ 1 \\
2
\end{gathered}\right) = \left(
\begin{gathered}
2\ \ 1\ \ 0\\
1\ \ 0 \\
0
\end{gathered}\right)\,,
\quad R_{g_2}\left(
\begin{gathered}
2\ \ 1\ \ 0\\
1\ \ 0  \\
0
\end{gathered}\right) = \left(
\begin{gathered}
2\ \ 1\ \ 0\\
2\ \ 1 \\
2
\end{gathered}\right)\,,
\quad R_{g_2}\left(
\begin{gathered}
2\ \ 1\ \ 0\\
2\ \ 1 \\
1
\end{gathered}\right) = -\left(
\begin{gathered}
2\ \ 1\ \ 0\\
1\ \ 0 \\
1
\end{gathered}\right)\,,
$$
$$
R_{g_2}\left(
\begin{gathered}
2\ \ 1\ \ 0\\
1\ \ 0 \\
1
\end{gathered}\right) = -\left(
\begin{gathered}
2\ \ 1\ \ 0\\
2\ \ 1 \\
1
\end{gathered}\right)\,,
\quad R_{g_2}\left(
\begin{gathered}
2\ \ 1\ \ 0\\
2\ \ 0 \\
2
\end{gathered}\right) = -\left(
\begin{gathered}
2\ \ 1\ \ 0\\
2\ \ 0 \\
0
\end{gathered}\right)\,,
\quad R_{g_2}\left(
\begin{gathered}
2\ \ 1\ \ 0\\
2\ \ 0 \\
2
\end{gathered}\right) =- \left(
\begin{gathered}
2\ \ 1\ \ 0\\
2\ \ 0 \\
2
\end{gathered}\right)\,,
$$
$$
R_{g_2}\left(
\begin{gathered}
2\ \ 1\ \ 0\\
1\ \ 1 \\
1
\end{gathered}\right) = \left(
\begin{gathered}
2\ \ 1\ \ 0\\
1\ \ 1 \\
1
\end{gathered}\right)\,,
\quad R_{g_2}\left(
\begin{gathered}
2\ \ 1\ \ 0\\
2\ \ 0 \\
1
\end{gathered}\right) = \left(
\begin{gathered}
2\ \ 1\ \ 0\\
2\ \ 0 \\
1
\end{gathered}\right)\,.
$$

If the representation $r=r(m_{13}, m_{23}, 0)$ is not
self-contragradient, then the grading $\Gamma_2$ is compatible
with the reducible representation
$$ r_\oplus(X) :=  \left(\begin{array}{cc} r(X)&0\\0&-\bigl(r(X)\bigr)^T \end{array} \right)
$$
 and the corresponding simulation matrix on the double-dimensional
 space is $J =\sigma_1\otimes I$, where $I$ is the identity operator on the
 representation space of  representation $r$.

%
%
%

%

\section*{Acknowledgements}
We are grateful to Vyacheslav Futorny for fruitful discussions on
relations between notions grading and group grading and to Ji\v
r\'{\i} Patera for introducing us to the problems connected with
representations of contracted  Lie algebras.  We acknowledge
financial support by the grants MSM6840770039 and LC06002  of the
Ministry of Education, Youth, and Sports of the Czech Republic.

\section*{Appendix}
Let us give explicit description of irreducible representation of
$sl(n,\mathbb{C})$ in Gel'fand Tseitlin formalism. Since any
$E_{k\ell}$ can be obtained by commutation relations from
$E_{k,k}$, $E_{k,k-1}$ and
 $E_{k-1,k}$, only formulas  for $r(E_{k,k})$,  $r(E_{k,k-1})$ and
 $r(E_{k-1,k})$ are needed.

$$r(E_{k,k})\xi({\bf m})= (r_{k}- r_{k-1})\xi({\bf m})\,,
$$
 where $r_k = m_{1,k}+ \ldots +  m_{k,k}$ \ for \  $k=1,2,\ldots,n$ and
 $r_0 = 0$,
$$r(E_{k,k-1})\xi({\bf m})= a_{k-1}^{1}\xi({\bf m}^{1}_{k-1})+\ldots +
 a_{k-1}^{k-1}\xi({\bf m}^{k-1}_{k-1})\,,
$$
where ${\bf m}^{j}_{k-1}$ denotes the triangular pattern obtained
from ${\bf m} $  replacing $m_{j,k-1}$  by   $m_{j,k-1} -1$,
$$ a_{k-1}^j = \left[ - \frac{~\prod_{i=1}^k(m_{ik}-m_{j,k-1}-i+j+1)
\prod_{i=1}^{k-2}(m_{i,k-2}-m_{j,k-1}-i+j)}{ ~\prod_{i\neq j}
(m_{i,k-1}-m_{j,k-1}-i+j+1)(m_{i,k-1}-m_{j,k-1}-i+j)
}\right]^{1/2}
$$
$$r(E_{k-1,k})\xi({\bf m})= b_{k-1}^{1}\xi({\bf m}^{1}_{k-1})+\ldots +
 b_{k-1}^{k-1}\xi({\bf m}^{k-1}_{k-1})\,,
$$
where ${\bf m}^{j}_{k-1}$ denotes the triangular pattern obtained
from ${\bf m} $  replacing $m_{j,k-1}$  by   $m_{j,k-1} +1$, and
$$ b_{k-1}^j = \left[ - \frac{~\prod_{i=1}^k(m_{ik}-m_{j,k-1}-i+j)
\prod_{i=1}^{k-2}(m_{i,k-2}-m_{j,k-1}-i+j-1)}{ ~\prod_{i\neq j}
(m_{i,k-1}-m_{j,k-1}-i+j)(m_{i,k-1}-m_{j,k-1}-i+j-1)
}\right]^{1/2}
$$

\end{document}